\documentclass[12pt] {article}
\usepackage[active]{srcltx}
\usepackage{latexsym}
\usepackage[dvips]{graphics,color}
\usepackage{graphpap}
\usepackage{alltt}
\usepackage{graphicx}
\usepackage{amsfonts}
\usepackage{amsmath}
\usepackage{psfrag}
\newtheorem{theorem}{Theorem}[section]
\newtheorem{lemma}[theorem]{Lemma}
\newtheorem{proposition}[theorem]{Proposition}

\newtheorem{definition}[theorem]{Definition}
\newtheorem{corollary}{Corollary}[section]
\newcommand{\beq}{\begin{equation}}
\newcommand{\feq}[1]{\label{#1} \end{equation}}
\newcommand{\beqr}{\begin{eqnarray}}
\newcommand{\feqr}{\end{eqnarray}}
\def\non{\nonumber}

\newcommand{\rf}[1]{(\ref{#1})}
\begin{document}
\title{\bf Spectral Properties of the Dirichlet Operator $\sum_{i=1}^{d}(-\partial_i^2)^{s}$ on Domains in d-Dimensional Euclidean Space\unboldmath}
\author{Agapitos N. Hatzinikitas, \\
University of Aegean, 
School of Sciences, \\
Department of Mathematics, 
Karlovasi 83200,
Samos, Greece \\
Email: ahatz@aegean.gr}
\date{}
%%%%%%%%%%%%%%%%%%%%%%%%%%%%%%%%%%%%%%%%%%%%%%%%%%%%%%%%%%%%%%%%%%%%%%%%%%%%%%%%%%%%%%%%%%%%%%%%%%
%%%%%%%%%%%%%%%%%%%%%%%%%%%%%%%%%%%%%%%%%%%%%%%%%%%%%%%%%%%%%%%%%%%%%%%%%%%%%%%%%%%%%%%%%%%%%%%%%%
\maketitle

\begin{abstract}
In this article we investigate the distribution of eigenvalues of the Dirichlet pseudo-differential operator $\sum_{i=1}^{d}(-\partial_i^2)^{s}, \, s\in (\frac{1}{2},1]$ on an open and bounded subdomain  $\Omega \subset \mathbb{R}^d$ and predict bounds on the sum of the first $N$ eigenvalues, the counting function, the Riesz means and the trace of the heat kernel. Moreover, utilizing the connection of coherent states to the semi-classical approach of Quantum Mechanics we determine the sum for moments of eigenvalues of the associated Schr\"{o}dinger operator.
\end{abstract}
%%%%%%%%%%%%%%%%%%%%%%%%%%%%%%%%%%%%%%%%%%%%%%%%%%%%%%%%%%%%%%%%%%%%%%%%%%%%%%%%%%%%%%%%%%%%%%%%%%%%%%%%%%%%%
%%%%%%%%%%%%%%%%%%%%%%%%%%%%%%%%%%%%%%%%%%%%%%%%%%%%%%%%%%%%%%%%%%%%%%%%%%%%%%%%%%%%%%%%%%%%%%%%%%%%%%%%%%%%%
\vspace{1cm}
\newpage
%%%%%%%%%%%%%%%%%%%%%%%%%%%%%%%%%%%%%%%%%%%%%%%%%%%%%%%%%%%%%%%%%%%%%%%%%%%%%%%%%%%%%%%%%%%%%%%%%%%%%%%%%%%
\section{Introduction}
In 1912 H. Weyl \cite{W1}, in a brilliant solution to the asymptotic behaviour of the sequence of eigenvalues for the Dirichlet Laplacian over the bounded domain $\Omega \subset \mathbb R^2$, proved that
\beqr
\lim_{k\rightarrow \infty}\frac{k}{\mathcal{E}_k}=\frac{|\Omega|}{4\pi}
\label{sec1 : eq1}
\feqr
where $|\Omega|$ is the surface area of $\Omega$. Defining the counting function $\mathcal{N}(\mathcal{E}):=\sharp  \{ \mathcal{E}_n \leq \mathcal{E} \} $, relation \rf{sec1 : eq1} is equivalent to 
\beqr
\mathcal{N}(\mathcal{E})=\frac{|\Omega|}{4\pi}\mathcal{E}+o(\mathcal{E}) \quad \textrm{as} \quad \mathcal{E}\rightarrow \infty.
\label{sec1 : eq2}
\feqr
These reasults are now called \textit{Wel's law}. Shortly afterwards, he submitted two papers \cite{W2,W3}  which contain a generalization of \rf{sec1 : eq2} to the three dimensional scalar wave equation and the extension to the vector Helmholtz wave equation describing the vibrations of the electric field $\vec{E}$ in an empty cavity $\Omega$ with perfectly reflecting walls $\partial \Omega$. Later he conjectured the existence of a second asymptotic term of lower order in two and three dimensions
\beqr
\mathcal{N}(\mathcal{E})=\left\{\begin{array}{lll} \frac{|\Omega|}{4\pi}\mathcal{E}\mp \frac{|\partial \Omega|}{4\pi}\sqrt{\mathcal{E}}+o(\sqrt{\mathcal{E}}),\quad & d=2,   \quad & \mathcal{E}\rightarrow \infty \\ \\
\frac{|\Omega|}{6\pi^2}\mathcal{E}^{\frac{3}{2}}\mp \frac{|\partial \Omega|}{16\pi}\mathcal{E}+o(\mathcal{E}), \quad& d=3, \quad & \mathcal{E}\rightarrow \infty  \end{array} \right.
\label{sec1 : eq3}
\feqr
where $|\partial \Omega|$ denotes the length of the circumference of the domain in $d=2$ and the surface area in $d=3$ dimensions respectively. Also the minus sign refers to the Dirichlet boundary condition $u|_{\partial \Omega}=0$ and the plus sign to the Neumann boundary condition $\partial u/\partial n =0, \, x\in \partial \Omega$. These formulae were justified (under a global condition on the geometry of $\Omega$) by V. Ivrii \cite{IV} and R. Melrose \cite{MEL} in 1980.
\par In 1959 R. Blumenthal and R. Getoor \cite{BG} obtained the following result 
\beqr
\mathcal{N}(\mathcal{E})=\frac{|\Omega|}{(4\pi)^{\frac{d}{2}}\Gamma(1+\frac{d}{2})} \mathcal{E}^{\frac{d}{2s}}+o(\mathcal{E}^{\frac{d}{2s}}), \, s\in (0,1], \quad \mathcal{E}\rightarrow \infty
\label{sec1 : eq4}
\feqr
for the asymptotic distribution of the eigenvalues for a symmetric stable process of index $\alpha$, with infinitesimal generator the fractional Laplacian $(-\Delta)^s|_{\Omega}$, by applying Karamata's Tauberian theorem. 
\par G. P\'{o}lya \cite{P} in 1961 conjectured for an arbitrary domain and proved only for tiling domains, i.e. domains whose congruent non-overlapping translations cover $\mathbb{R}^d$ without gaps, that
\beqr
\mathcal{N}(\mathcal{E})\leq \frac{|\Omega|}{(2\pi)^d}\left(\frac{|S_{d-1}|}{d}\right)\mathcal{E}^{\frac{d}{2}}.
\label{sec1 : eq5}
\feqr 
For general domains the conjecture is still open although extensions to product domains $\Omega_1\times \Omega_2\subset \mathbb R^{d_1+d_2}$, where $\Omega_1\subset \mathbb R^{d_1}, \, d_1\geq 2$ is a tiling domain and $\Omega_2\subset \mathbb R^{d_2}, \, d_2\geq 1$ is an arbitrary domain of finite Lebesgue measure, can be found in \cite{La}.
\par The closest result to P\'{o}lya's inequality for an arbitrary bounded domain in $\mathbb R^d$ is due to F. Berezin \cite{BE} and, P. Li and S. Yau \cite{LY} who proved the sharp bound
\beqr
\sum_{n=1}^{k}\mathcal{E}_n\geq \frac{d}{d+2} \frac{4\pi^2}{(|B_d| |\Omega|)^{\frac{2}{d}}} k^{1+\frac{2}{d}}, \quad k\in \mathbb N, \quad |B_d|=\frac{1}{d}|S_{d-1}|
\label{sec1 : eq6}
\feqr
where $|B_d|$ is the volume of the d-ball.
\par This paper is structured as follows:
\par In Sec.2 we provide the definition of the $\hbar-$dependent unitary Fourier transform operator as well as that of the operator $\sum_{i=1}^{d}(-\partial_i^2)^{s}$ through relation \rf{sec2 : eq1a2}. The discreteness of the spectrum for the Dirichlet problem on an open and bounded $d-$dimensional hypercube with the assistance of Lemma \rf{sec2 : eq5d} enable us to prove Weyl's law. Next we estimate the sum of the first $N$ eigenvalues and by Theorem \rf{sec2 : eq14} we prove P\'{o}lya's inequality for a tiling domain. As a corollary we derive the lower bound of the aforementioned sum. Theorem \rf{sec2 : eq14ha} generalizes P\'{o}lya's inequality for an open, bounded and simply connected subdomain of $\mathbb R^d$ and also predicts an upper bound for the counting function. 
\par In Sec.3 we estimate the Riesz's mean of order $\rho\geq 0$ and the partition function taking advantage of their interconnection through a Laplace transform. Upper bounds for both quantities are also found for $\rho>1$. 
\par In Sec.4 considering a particle moving freely in a subset of the phase space and using the semi-classical approximation method we determine the sum of its eigenvalues. The result coincides to the one derived from Weyl's law or P\'{o}lya's inequality after an appropriate rescaling of the tiling domain. In the presence of a negatively valued potential $V\in L^{1+\frac{d}{2s}}(\mathbb{R}^d)$ by performing a similar calculation we exctract \rf{sec3 : eq5} for the sum of the eigenvalues. 
\par In Sec.5 we begin with the definition and basic properties of coherent states. In the sequel, by making a suitable choice for the normalized coherent states, we find  the semi-classical limit of the expectation value for the corresponding Schr\"{o}dinger operator. Finally, Theorem \rf{sec4 : eq9a1} establishes the  semi-classical sum for moments of eigenvalues of the Schr\"{o}dinger operator.           
%%%%%%%%%%%%%%%%%%%%%%%%%%%%%%%%%%%%%%%%%%%%%%%%%%%%%%%%%%%%%%%%%%%%%%%%%%%%%%%%%%%%%%%%%%%%%%%%%%%%%%%%%%%%%
%%%%%%%%%%%%%%%%%%%%%%%%%%%%%%%%%%%%%%%%%%%%%%%%%%%%%%%%%%%%%%%%%%%%%%%%%%%%%%%%%%%%%%%%%%%%%%%%%%%%%%%%%%%%%
\section{The counting function and the sum of eigenvalues for the Dirichlet problem}

\begin{definition}
\label{sec2 : eq1a1}
Given $\psi \in S(\mathbb{R}^d)$, where $S$ is the Schwartz space\footnote{The linear space consisting of all $\psi\in C^{\infty}(\mathbb{R}^d)$ for which \begin{displaymath} 
\textrm{sup}_{x\in \mathbb{R}^d}|x^n (D^m)\psi)(x)|<\infty.\end{displaymath}},  we denote by $\mathcal{F}_{\hbar}$ the unitary $\hbar-$dependent Fourier operator 
\beqr
\mathcal{F}_{\hbar} : S(\mathbb{R}^d) \rightarrow S(\mathbb{R}^d)
\label{sec2 : eq1a1a}
\feqr
defined by
\beqr
(\mathcal{F}_{\hbar}\psi)(p)=\hat{\psi}(p)=\frac{1}{(2\pi \hbar)^{\frac{d}{2}}}\int_{\mathbb{R}^d} e^{-\frac{i}{\hbar}\langle p,x\rangle}\psi(x)dx.  
\label{sec2 : eq1a2} 
\feqr
\end{definition}
The integral in \rf{sec2 : eq1a2} is understood as the limit $\hat{\psi}=\lim_{n\rightarrow \infty} \hat{\psi}_n$ in the strong topology in $L^2(\mathbb{R}^d)$, where
\begin{displaymath}
\hat{\psi}_n(p)=\frac{1}{(2\pi \hbar)^{\frac{d}{2}}}\int_{-n}^{n} e^{-\frac{i}{\hbar}\langle p,x\rangle}\psi(x)dx, \quad n\in \mathbb{R}^d.   
\end{displaymath}
The inverse Fourier transformation is given by
\beqr
(\mathcal{F}_{\hbar}^{-1}\hat{\psi})(x)=\psi (x)=\frac{1}{(2\pi \hbar)^{\frac{d}{2}}}\int_{\mathbb{R}^d} e^{\frac{i}{\hbar}\langle p,x\rangle}\hat{\psi}(p)dp.
\label{sec2 : eq1a3}
\feqr
\begin{definition}
\label{sec2 : eq1a}
Let $s\in (\frac{1}{2},1]$, $\psi \, : \mathbb{R}^d \rightarrow \mathbb{R}$ and  
\begin{displaymath}
\mathcal{L}_{2s,\hbar} \, : S(\mathbb{R}^d) \rightarrow L^2(\mathbb{R}^d) 
\end{displaymath}
where $\mathcal{L}_{2s, \hbar}=-\sum_{i=1}^{d}(-\hbar^2\partial_i^2)^{s}$ \footnote{The $\hbar$ dependence of the operator $\mathcal{L}$ will be declared explicitly when needed.} and $\partial_i$ denotes  the partial derivative w.r.t. $x_i$. We define the operator $\mathcal{L}_{2s,\hbar}$ by 
\begin{eqnarray}
(\mathcal{L}_{2s, \hbar}\psi)(x)&:=&\frac{1}{(2\pi \hbar)^{\frac{d}{2}}}\int_{{\mathbb R^d}}e^{\frac{i}{\hbar}\langle p,x\rangle}\left\| p\right\|^{2s} \hat{\psi}(p)dp \non \\
&=& \left(\mathcal{F}_{\hbar}^{-1}\hat{g}\right)(x), \quad \hat{g}(p)=\left\| p\right\|^{2s}(\mathcal{F}_{\hbar} \psi)(p).
\label{sec2 : eq1}
\end{eqnarray}
In \rf{sec2 : eq1} $\hbar$ is Planck's constant and $\left\| p \right\|^{2s}=\sum_{i=1}^{d}|p_i|^{2s}$ is the $2s-$norm\footnote{The Euclidean norm will be denoted by $\left\|\cdot \right\|^2_2$.} corresponding to the symbol of the pseudo-differential operator \cite{S}.
\end{definition}
Note that $(-\Delta)^s=(-\sum_{i=1}^d \partial_i^2)^s \neq \sum_{i=1}^{d}(-\partial_i^2)^{s}$ unless $s=1$. Definition \rf{sec2 : eq1a} is initiated by the anisotropic fractional diffusion equation 
\beqr
\frac{\partial \psi(x,t)}{\partial t}=-\sum_{i=1}^d D_i (-\partial_i^2)^{s} \psi(x,t), \quad (x,t)\in \mathbb{R}^d \times [0,\infty]
\label{sec2 : eq2}
\feqr
considered in \cite{BM}. 
\begin{proposition}
On the open and bounded hypercube $\Gamma_d\subset {\mathbb R^d}$, the eigenvalues for the homogeneous Dirichlet problem
\beqr
\left(\sum_{i=1}^{d}(-\partial_i^2)^s \psi_n \right)(x) &=&\mathcal{E}_n\psi_n(x), \,\, \textrm{in} \,\, \Gamma_d; \,\,\mathcal{E}_n=\frac{E_n}{D_{2s}} \non \\
 \psi_n(x)&=& 0 \,\,  \textrm{on} \,\, \overline{\Gamma}_d
\label{sec2 : eq4}
\feqr
are given by 
\begin{eqnarray}
\mathcal{E}_{n} =\left\| \frac{n \pi}{L}\right\|^{2s}, \quad n\in {\mathbb Z_+^d} 
\label{sec2 : eq5}
\end{eqnarray}
 where $\{\psi_n\}_{n=1}^{\infty}$ forms an orthonormal basis in $L^2(\Gamma_d)$ with $\psi_n(x)=c_n\prod_{j=1}^{d}\psi_{n_j}(x_j)$, at least one $n_j$ should not vanish, and $D_{2s}$ is a constant with dimensions $[D_{2s}]=[M]^{1-2s}\left([L]/[T]\right)^{2(1-s)}$.
\end{proposition}
\textbf{Proof.} The Fourier transformation of the boundary conditions requires $p_j$'s to be discrete and moreover applying Parseval's identity to \rf{sec2 : eq4}  it can be proved that the eigenvalues $\mathcal{E}_n$ should also be discrete and given by \rf{sec2 : eq5} provided one makes the substitution $p_j=n_j \pi/L$. \hfill\(\Box\) \\
Arranging the positive, real and discrete spectrum of $\mathcal{L}_{2s}$ in increasing order (including multiplicities), we have 
\beqr
0<\mathcal{E}_1(\Gamma_d) < \mathcal{E}_2(\Gamma_d) < \mathcal{E}_3(\Gamma_d) < \cdots \quad \textrm{and} \,\, \lim_{n\rightarrow \infty} \mathcal{E}_n(\Gamma_d) =\infty.
\label{sec2 : eq5a}
\feqr
\textbf{Remark.} If $\Gamma'_d\subset \Gamma_d \subset \mathbb R^d$ such that $|\Gamma'_d|=\lambda^d|\Gamma_d|$ where the scale factor $\lambda\in(0,1)$ then the nth eigenvalues satisfy
\beqr
\mathcal{E}_n(\Gamma_d)=\lambda^{2s}\mathcal{E}_n(\Gamma'_d)
\label{sec2 : eq5b}
\feqr
as can be checked by \rf{sec2 : eq5}. The scaling property \rf{sec2 : eq5b} can be generalized as follows: let $\Omega'\subset \Omega \subset \mathbb R^d$ and $|\Omega'|=\lambda^d|\Omega|$ then
\beqr
\mathcal{E}_n(\Omega)=\lambda^{2s}\mathcal{E}_n(\Omega').
\label{sec2 : eq5c}
\feqr 
This statement can be proved using \rf{sec2 : eq1} with $\hat{g}(p)=\left\| p\right\|^{2s}(\mathcal{F}_{\hbar} (\chi_{\Omega}\psi))(p)$ and making the change of variables $x=\lambda z$.
\\ 
The following Lemma \cite{KO} will be used repeatedly in our study. 
\begin{lemma}
\label{sec2 : eq5d}
The integral formula 
\begin{equation}
\int_{{\mathbb R^d}}e^{-\left\| x\right\|^{2s}} dx=\left(2\Gamma\left(1+\frac{1}{2s}\right)\right)^d
\label{sec2 : eq6}
\end{equation}
 holds and one may recover from it the volume 
\begin{equation}
|B_{d,2s}|:=\textrm{Vol} (B_{d,2s})=\frac{\left(2\Gamma\left(1+\frac{1}{2s}\right)\right)^d}{\Gamma\left(1+\frac{d}{2s}\right)}
\label{sec2 : eq7}
\end{equation}
of the convex unit ball defined as 
\begin{displaymath}
B_{d,2s}=\biggr\{x\in {\mathbb R^d}: \left\| x\right\|=\left(\sum_{i=1}^n |x_i|^{2s}\right)^{\frac{1}{2s}}\leq 1\biggl\}. 
\end{displaymath}
\end{lemma}
\textbf{Proof.} 
Starting from the left-hand side of (\ref{sec2 : eq6}) we have
\begin{eqnarray}
\int_{{\mathbb R^d}}e^{-\left\| x\right\|^{2s}} dx &=& 2^d \prod_{i=1}^{d}\left(\int_{0}^{\infty} e^{-|x_i|^{2s}} dx_i\right) 
= \left(\frac{1}{s}\int_{0}^{\infty}u^{\frac{1}{2s}-1} e^{-u} du\right)^d \non \\
&=& \left(2 \Gamma\left(1+\frac{1}{2s}\right)\right)^d
\label{sec2 : eq8} 
\end{eqnarray}
where the factor $2^d$ represents the number of orthants and the gamma function $\Gamma$ is defined by Euler's integral of the second kind 
\beqr
\Gamma(z)=\int_{0}^{\infty}e^{-t}t^{z-1}dt, \, \textrm{Re z}>0. 
\label{sec2 : eq8a}
\feqr
On the other hand, the same integral can be computed as 
\begin{eqnarray}
\int_{{\mathbb R^d}}e^{-\left\| x\right\|^{2s}} dx \!\!\!&=&\!\!\!\! \int_{{\mathbb R^d}}\left(\int_{\left\| x\right\|^{2s}}^{\infty}e^{-u}du\right) dx 
=\!\! \int_{0}^{\infty} \!\!e^{-u} du \left(\int_{{\mathbb R^d}}\chi(\{u\in [0,\infty): \left\| x\right\|\leq u^{\frac{1}{2s}}\}) dx\right) \non \\
\!\!\!&=&\!\!\! \int_{0}^{\infty} e^{-u} |u^{\frac{1}{2s}}B_{d,2s}|du 
= |B_{d,2s}| \int_{0}^{\infty} u^{\left(1+\frac{d}{2s}\right)-1} e^{-u} du \non \\
\!\!\!&=&\!\!\! |B_{d,2s}| \Gamma\left(1+\frac{d}{2s}\right).
\label{sec2 : eq9} 
\end{eqnarray}
Comparing the two expressions \rf{sec2 : eq8} and \rf{sec2 : eq9} we get the result \rf{sec2 : eq7}.  \hfill\(\Box\) \\
\textbf{Remark.} If one uses the Euclidean norm $\left\| \cdot \right\|_{2}^{2}$ then \rf{sec2 : eq7} becomes
\beqr
|B_{d,2s}|_E=\frac{1}{d}|S_{d-1}|, \quad |S_{d-1}|=\frac{2\pi^{\frac{d}{2}}}{\Gamma(\frac{d}{2})}. 
\label{sec2 : eq9a}
\feqr
\begin{proposition}
The number of eigenvalues $\mathcal{E}_n$ in a d-dimensional, 2s-deformed hypersphere of radius $R=\frac{L}{\pi}\mathcal{E}^{\frac{1}{2s}}, \, s\in (\frac{1}{2},1]$ asymptotically ($R\rightarrow \infty$) is given by the counting function
\begin{equation}
\mathcal{N}(\mathcal{E})=\frac{|\Gamma_d|\mathcal{E}^{\frac{d}{2s}}}{(2\pi)^d d}|A_{d-1,2s}|+ o(\mathcal{E}^{\frac{d}{2s}}), \quad |A_{d-1,2s}|=\frac{2s \left(2\Gamma(1+\frac{1}{2s})\right)^d}{\Gamma\left(\frac{d}{2s}\right)}=d|B_{d,2s}|
\label{sec2 : eq10a} 
\end{equation}
 where $|A_{d-1,2s}|$ represents the volume of the 2s-deformed unit sphere $S_{d-1}$ and the little $o(\cdot)$ symbol means a term that grows slower than $(\cdot)$.
\end{proposition}
\textbf{Proof.}
 Using the definition of the counting function (i.e. the function that counts the number of eigenvalues not exceeding a cut off value $\mathcal{E}$) we have
\begin{eqnarray}
\mathcal{N}(\mathcal{E}) &:=& \sum_{\mathcal{E}_n\leq \mathcal{E}} 1= \sharp  \{ n\in {\mathbb Z^d_+}: \mathcal{E}_n \leq \mathcal{E} \} \non \\
&\stackrel{\rf{sec2 : eq5}}{=}& \sharp \{ n\in {\mathbb Z^d_+}: \sum_{i=1}^{d} |n_i|^{2s}\leq \left(\frac{L}{\pi} \mathcal{E}^{\frac{1}{2s}}\right)=R\} \non \\
&=& \frac{1}{2^d} R^d |B_{d,2s}| + o(\mathcal{E}^{\frac{d}{2s}}), \quad R\rightarrow \infty \non \\
&=& \frac{1}{(2\pi)^d}\frac{|\Gamma_d||A_{d-1,2s}|}{d}\mathcal{E}^{\frac{d}{2s}}+ o(\mathcal{E}^{\frac{d}{2s}}), \quad \mathcal{E}\rightarrow \infty
\label{sec2 : eq10b} 
\end{eqnarray}
where Lemma (\ref{sec2 : eq7}) has been applied. \hfill\(\Box\)\\

\textbf{Remarks.}
\begin{enumerate}
\item Solving (\ref{sec2 : eq10a}) w.r.t. $\mathcal{E}:=\mathcal{E}_{N}$ we obtain 
\begin{equation}
\mathcal{E}_{N}= (2\pi)^{2s}\left( \frac{\mathcal{N}d}{|A_{d-1,2s}||\Gamma_d|}\right)^{\frac{2s}{d}}+o(\mathcal{N}^{\frac{2s}{d}}).
\label{sec2 : eq11} 
\end{equation}  
Relation \rf{sec2 : eq11} represents an extension of Blumenthal's and Getoor's result which in the Euclidean norm case is given by \rf{sec1 : eq4}.
Summing the eigenvalues (\ref{sec2 : eq11}) using the finite series formula \cite{GR}
\begin{displaymath} 
\sum_{k=1}^{n}k^q=\frac{n^{q+1}}{q+1}+\frac{n^q}{2}+o(n^q) 
\end{displaymath}
we have
\begin{equation}
 S(N):=\sum_{n=1}^{N} \mathcal{E}_n= (2\pi)^{2s}\frac{d}{d+2s}\left( \frac{d}{|A_{d-1,2s}||\Gamma_d|}\right)^{\frac{2s}{d}}\mathcal{N}^{1+\frac{2s}{d}}+o(\mathcal{N}^{1+\frac{2s}{d}}).
\label{sec2 : eq13}
\end{equation}
\item[($ii$)] Substituting the values $s=1$ and $d=2$ into relation \rf{sec2 : eq10a} we recover Weyl's asymptotic formula \rf{sec1 : eq2} for a square
while for the same values of $s,d$ into \rf{sec2 : eq11}  we confirm P\'{o}lya's result \rf{sec1 : eq5}.   
\end{enumerate}
\begin{theorem}[P\'{o}lya's inequality for $\mathcal{L}_{2s}$ over tiling domains]
\label{sec2 : eq14}
If $\Omega\subset \mathbb R^d$ is a tiling domain then
\beqr
\mathcal{E}_n\geq (2\pi)^{2s}\left( \frac{nd}{|A_{d-1,2s}||\Omega|}\right)^{\frac{2s}{d}}.
\label{sec2 : eq14a}
\feqr 
\end{theorem}
\textbf{Proof.} Let $\Omega'$ be another tiling subdomain of $\Omega$ such that $|\Omega'|=\lambda^d|\Omega|$ then \rf{sec2 : eq5c} holds. Also suppose $\Gamma^1_d$ is the unit hypercube in $\mathbb R^d$ and $m$ be the number of congruent domains $\Omega'$ filling $\Gamma^1_d$ without overlapping and leaving gaps. Then we obtain the following two relations
\beqr
\lim_{m\rightarrow \infty}( m|\Omega'|)&=&|\Gamma^1_d|=1\Rightarrow \lim_{m\rightarrow \infty} (m \lambda^{d})=\frac{1}{|\Omega|} \quad \textrm{and} \label{sec2 : eq14b} \\
\mathcal{E}''_{nm}(\Gamma^1_d)&\leq& \mathcal{E}'_n(\Omega')=\frac{1}{\lambda^{2s}}\mathcal{E}_n(\Omega) \label{sec2 : eq14c}
\feqr
where by $\mathcal{E}(\cdot)$ we denote the eigenvalue on the corresponding domain. By virtue of \rf{sec2 : eq11} and combining \rf{sec2 : eq14b}, \rf{sec2 : eq14c} we have
\beqr
\mathcal{E}_n(\Omega) \geq \frac{\mathcal{E}''_{nm}(\Gamma^1_d)}{(nm)^{\frac{2s}{d}}} (nm)^{\frac{2s}{d}}\lambda^{2s}.
\label{sec2 : eq14d}
\feqr
Taking the $m\rightarrow \infty$ limit we finally find
\beqr
\mathcal{E}_n(\Omega) \geq (2\pi)^{2s}\left( \frac{nd}{|A_{d-1,2s}||\Omega|}\right)^{\frac{2s}{d}}. 
\label{sec2 : eq14e}
\feqr  
\hfill\(\Box\)
\begin{corollary}
 If $\Omega \subset \mathbb R^d$ is a tiling domain then
\beqr
S(N)\geq (2\pi)^{2s}\frac{d}{d+2s}\left( \frac{d}{|A_{d-1,2s}||\Gamma_d|}\right)^{\frac{2s}{d}}\mathcal{N}^{1+\frac{2s}{d}}.
\label{sec2 : eq14f}
\feqr
\end{corollary}
\textbf{Proof.} The function $f(t)=t^{\frac{2s}{d}}$ is increasing for $t\geq 0$ and applying the inequality
\beqr
\sum_{n=0}^{N-1}f(n) \leq \int_{0}^{N}f(t)dt\leq \sum_{n=0}^{N-1}f(n+1)
\label{sec2 : eq14g}
\feqr
with the help of \rf{sec2 : eq14f}, we show that
\beqr
S(N) &\geq& (2\pi)^{2s}\left( \frac{d}{|A_{d-1,2s}||\Omega|}\right)^{\frac{2s}{d}} \sum_{n=1}^{N-1} n^{\frac{2s}{d}}
\geq (2\pi)^{2s}\left( \frac{d}{|A_{d-1,2s}||\Omega|}\right)^{\frac{2s}{d}}  \int_{0}^{N} t^{\frac{2s}{d}} dt \non \\
&=& (2\pi)^{2s}\frac{d}{d+2s}\left( \frac{d}{|A_{d-1,2s}||\Omega|}\right)^{\frac{2s}{d}} N^{1+\frac{2s}{d}}. 
\label{sec2 : eq14h}
\feqr
\hfill\(\Box\)
\begin{theorem}
\label{sec2 : eq14ha}
Let $\Omega$ be an open, bounded and simply connected set in $\mathbb R^d$ of finite volume $|\Omega|$. Consider the homogeneous Dirichlet eigenvalue problem
\beqr
\left(\sum_{i=1}^{d}(-\partial_i^2)^s \psi_n \right)(x) &=&\mathcal{E}_n\psi_n(x), \quad in \,\, \Omega; \,\,\mathcal{E}_n=\frac{E_n}{D_{2s}} \non \\
 \psi_n(x)&=& 0 \,\,  on \,\, \overline{\Omega} \non \\
\langle \psi_n,\psi_m \rangle=\int_{\Omega} \bar{\psi}_n(x) \psi_m(x) dx&=&\delta_{mn}, \quad \forall m,n.
\label{sec2 : eq15} 
\feqr
Then
\beqr
S(N)\geq (2\pi)^{2s}\frac{d}{d+2s}\left( \frac{d}{|A_{d-1,2s}||\Omega|}\right)^{\frac{2s}{d}}N^{1+\frac{2s}{d}} 
\label{sec2 : eq16}
\feqr
and the following bound for the counting function valids 
\beqr
\mathcal{N}(z)\leq \frac{1}{(2\pi)^d}\left(\frac{d+2s}{d}\right)^{\frac{d}{2s}} \frac{|A_{d-1,2s}||\Omega|}{d} z^{\frac{d}{2s}}.
\label{sec2 : eq16a}
\feqr
\end{theorem}
\textbf{Proof.} Consider the extension of $\psi_n$'s by setting them identically zero outside their support, namely
\beqr
\phi_n(x)=\left\{\begin{array}{ll} \psi_n(x), & x\in \Omega \\ 0, &  x\in \bar{\Omega}.  \end{array}\right.
\label{sec2 : eq17}
\feqr
Define the function 
\beqr
F_N(p):=\sum_{n=1}^{N}|\hat{\phi}_n(p)|^2
\label{sec2 : eq18}
\feqr
and by Plancherel's theorem observe that
\beqr
\int_{\mathbb R^d} F_N(p) dp =\sum_{n=1}^{N} \int_{\mathbb R^d} |\hat{\phi}_n(p)|^2 dp=\sum_{n=1}^{N} \int_{\Omega} |\phi_n(x)|^2 dx 
= \sum_{n=1}^{N} 1=\mathcal{N}(\Omega).
\label{sec2 : eq19}
\feqr
Furthermore, using \rf{sec2 : eq1} and \rf{sec2 : eq15} we derive the expression
\beqr
\int_{\mathbb R^d} \left\|p\right\|^{2s} F_N(p) dp =\sum_{n=1}^{N} \int_{\mathbb R^d} \left\|p\right\|^{2s} |\hat{\phi}_n(p)|^2 dp 
=\sum_{n=1}^{N} \langle \phi_n, \mathcal{L}_{2s} \phi_n\rangle=\sum_{n=1}^{N} \mathcal{E}_n=S(N).
\label{sec2 : eq20}
\feqr
For every fixed $p \in \mathbb R^d$, since $exp(i\langle p,x \rangle)\in L^2(\Omega)$, it follows that 
\beqr
e^{i\langle p,x\rangle}=\sum_{m=1}^{\infty}c_m(p)\phi_m(x), \quad with \quad c_m(p)=\int_{\Omega}\phi_m(x)e^{i\langle p,x\rangle}dx.
\label{sec2 : eq21}
\feqr
Thus from \rf{sec2 : eq18} we deduce
\beqr
F_N(p)\leq \sum_{n=1}^{\infty}|\hat{\phi}_n(p)|^2=\frac{1}{(2\pi)^d}\left|\sum_{m=1}^{\infty}c_m(p) \right|^2=\frac{1}{(2\pi)^d}\int_{\Omega}dx=\frac{|\Omega|}{(2\pi)^d}
\label{sec2 : eq22}
\feqr
which is the $L^2-$norm of $exp(i\langle p,x \rangle)$. The function $F_{N, min}(p)$ that minimizes expression \rf{sec2 : eq20} and satisfies \rf{sec2 : eq19} and \rf{sec2 : eq22} should have the form
\beqr
F_{N, min}(p)=\frac{1}{(2\pi)^d}|\Omega|\chi(\mathcal{B}(0,r))
\label{sec2 : eq23}
\feqr
where $\mathcal{B}(0,r)$ is the $2s$-deformed ball with radius $r$ obeying 
\beqr
r^d=\frac{dN}{|\Omega||A_{d-1,2s}|}.
\label{sec2 : eq24}
\feqr
Plugging \rf{sec2 : eq23} into \rf{sec2 : eq20} with $p=2\pi k$ we arrive at the desired result.  
\par To prove \rf{sec2 : eq16a} we choose $z\in[\mathcal{E}_k,\mathcal{E}_{k+1}]$ and using \rf{sec2 : eq16} we have
\beqr
k\mathcal{E}_k \geq S(k)\geq (2\pi)^{2s}\frac{d}{d+2s}\left( \frac{d}{|A_{d-1,2s}||\Omega|}\right)^{\frac{2s}{d}}k^{1+\frac{2s}{d}}. 
\label{sec2 : eq25}
\feqr
But $k=\mathcal{N}(z)$ so
\beqr
z \geq \mathcal{E}_k \geq  (2\pi)^{2s}\frac{d}{d+2s}\left( \frac{d}{|A_{d-1,2s}||\Omega|}\right)^{\frac{2s}{d}}\mathcal{N}^{\frac{2s}{d}}
\label{sec2 : eq26}
\feqr
from which \rf{sec2 : eq16a} follows.
\hfill\(\Box\)\\
\textbf{Remark.} Relation \rf{sec2 : eq16} for $s=1$ is in agreement with Li's and Yau's result \rf{sec1 : eq6}. In terms of the counting function we obtain the upper bound
\beqr
\mathcal{N}(z)\leq \frac{1}{(4\pi)^{\frac{d}{2}}} \left(\frac{d+2}{d}\right)^{\frac{d}{2}}\frac{|\Omega|}{\Gamma\left(1+\frac{d}{2}\right)} z^{\frac{d}{2}}.
\label{sec2 : eq27}
\feqr
%%%%%%%%%%%%%%%%%%%%%%%%%%%%%%%%%%%%%%%%%%%%%%%%%%%%%%%%%%%%%%%%%%%%%%%%%%%%%%%%%%%%%%%%%%%%%%%%%%%%%%%
%%%%%%%%%%%%%%%%%%%%%%%%%%%%%%%%%%%%%%%%%%%%%%%%%%%%%%%%%%%%%%%%%%%%%%%%%%%%%%%%%%%%%%%%%%%%%%%%%%%%%%
\section{Riesz means and the partition function}
It is generally believed that things get more manageable if one considers averaged or smoothed versions of the counting function such as the Riesz mean or the trace of the heat kernel, the so-called partition function. 
\begin{definition}
The Riesz mean of order $\rho\geq 0$ is defined for $\mathcal{E}> 0$ by
\label{sec2a :eq1}
\beqr
R_{\rho}(\mathcal{E}):=\textrm{Tr}\left(\sum_{i=1}^{d}(-\partial_i^2)^{s}_{\Omega}-\mathcal{E}\right)^{\rho}_{-}=\sum_{j}(\mathcal{E}-\mathcal{E}_j)_+^{\rho}
\label{sec2a :eq2}
\feqr 
where $x_{\pm}:=(|x|\pm x)/2$ denotes the positive and negative part of $x\in \mathbb{R}$ respectively.
\end{definition}
The Riesz mean reduces to the counting function when $\rho\rightarrow  0^+$ while for $\rho\rightarrow  1^-$ is directly realated to the sum of eigenvalues. This quantity describes the energy of non-interacting fermionic particles trapped in $\Omega$ and plays an important role in physical applications. If $\mathcal{E}_j$ is considered to be a continuous variable then \rf{sec2a :eq2} is replaced by  
\beqr
R_{\rho}(\mathcal{E})=\int_{0}^{\infty}(\mathcal{E}-t)_+^{\rho} dN(t)=\rho \int_{0}^{\infty}(\mathcal{E}-t)_+^{\rho-1} N(t)dt. 
\label{sec2a :eq3}
\feqr  
Relation \rf{sec2a :eq3} is a limiting case of the following iteration property \cite{AL} 
\beqr
R_{\rho+\delta}(\mathcal{E})=\frac{1}{B(1+\rho,\delta)}\int_{0}^{\infty}(\mathcal{E}-t)_+^{\delta-1} R_{\rho}(t)dt, \quad \rho\geq 0, \delta>0
\label{sec2a :eq4}
\feqr
where$B(x,y)$ denotes the beta function defined by the functional relation
\beqr
B(x,y)=\frac{\Gamma(x)\Gamma(y)}{\Gamma(x+y)}. 
\label{sec2a :eq4a}
\feqr
We point out that \rf{sec2a :eq4} is nothing but a Riemann-Liouville fractional integral transform. Substituting \rf{sec2 : eq10b} into \rf{sec2a :eq3} for a tiling domain $\Omega \subset \mathbb R^d$ we learn that
\beqr
R_{\rho}(\mathcal{E})\sim L_{\rho, d}^{cl.} |\Omega| \mathcal{E}^{\rho+\frac{d}{2s}}\quad \textrm{as} \quad \mathcal{E}\rightarrow \infty 
\label{sec2a :eq5}
\feqr
where the classical constant is given by 
\beqr
\label{sec2a :eq6}
\quad L_{\rho, d}^{cl.}=\frac{1}{\pi^d}\frac{\Gamma^d(1+\frac{1}{2s})\Gamma(1+\rho)}{\Gamma(1+\rho+\frac{d}{2s})}.
\feqr
One can smooth the counting function even further and consider the partition function defined by
\beqr
\label{sec2a :eq7}
Z(t):=\textrm{Tr}\left(e^{\sum_{i=1}^{d}(-\partial_i^2)^{s}_{\Omega} t}\right)=\sum_{j=1}^{\infty}e^{-\mathcal{E}_j(\Omega) t}.
\feqr
If $\mathcal{E}_j$ is a continuous variable then \rf{sec2a :eq7} is written as \cite{HH}
\beqr
\label{sec2a :eq8}
Z(t)=\int_{0}^{\infty}e^{-\mathcal{E}t}dN(\mathcal{E})=t\int_{0}^{\infty}e^{-\mathcal{E}t}N(\mathcal{E})d\mathcal{E}=tL[N(\cdot)](t) 
\feqr
where $\quad L[f(\cdot)](t)=\int_{0}^{\infty} e^{-zt}f(z)dz$ is the Laplace transform of a suitable function $f : (0,\infty) \rightarrow \mathbb{R}$. Again using  \rf{sec2 : eq10b} into \rf{sec2a :eq8} we have
\beqr
\label{sec2a :eq9}
Z(t)\sim \frac{1}{(2\pi)^d}|\Omega|\frac{(2\Gamma(1+\frac{1}{2s}))^d}{t^{\frac{d}{2s}}}
\feqr  
where $L[z^{\delta}](t)=\frac{\Gamma(1+\delta)}{t^{1+\delta}}$. \\

\textbf{Remarks.}
\begin{enumerate}
\item Utilizing Theorem \rf{sec2 : eq14ha} and applying the Laplace transform to inequality \rf{sec2 : eq16a}, it follows immediately that
\beqr
Z(t)\leq \frac{1}{(2\pi)^d}\left(\frac{d+2s}{d}\right)^{\frac{d}{2s}} \frac{|A_{d-1,2s}||\Omega|}{d} \Gamma\left(1+\frac{d}{2s}\right)t^{-\frac{d}{2s}}.
\label{sec2a :eq10}
\feqr 
\item The Laplace transform of \rf{sec2a :eq2} for $\rho>1$ and definition \rf{sec2 : eq8a} of gamma function leads to
\beqr
L[R_{\rho}(\cdot)](t)=\frac{\Gamma(1+\rho)}{t^{1+\rho}}\sum_j e^{-\mathcal{E}_j t}=\frac{\Gamma(1+\rho)}{t^{1+\rho}}Z(t).
\label{sec2a :eq11}
\feqr
Combining \rf{sec2a :eq10} with \rf{sec2a :eq11} we obtain the inequality
\beqr
R_{\rho}(\mathcal{E})\leq \frac{1}{(2\pi)^d}\left(\frac{d+2s}{d}\right)^{\frac{d}{2s}} \frac{|A_{d-1,2s}||\Omega|}{d} \rho B\left(\rho, 1+\frac{d}{2s}\right) \mathcal{E}^{\rho+\frac{d}{2s}}.
\label{sec2a :eq12}
\feqr
\end{enumerate}

%%%%%%%%%%%%%%%%%%%%%%%%%%%%%%%%%%%%%%%%%%%%%%%%%%%%%%%%%%%%%%%%%%%%%%%%%%%%%%%%%%%%%%%%%%%%%%%%%%%%%%%
%%%%%%%%%%%%%%%%%%%%%%%%%%%%%%%%%%%%%%%%%%%%%%%%%%%%%%%%%%%%%%%%%%%%%%%%%%%%%%%%%%%%%%%%%%%%%%%%%%%%%%

\section{Phase space and the semi-classical approximation for the sum of eigenvalues}

An alternative way \cite{LL} to reproduce (\ref{sec2 : eq13}) is to consider a classical particle moving freely inside a simply connected open subset $\Omega$ of ${\mathbb R^d}$ with reflective boundary. 
The state of the particle at any time is described by the $2n$-tuple $(x^1,\cdots,x^d,p_1,\cdots,p_d)$ of positions and momenta. The set of all allowed pairs $(\textbf{x},\textbf{p})$ is called phase space and it will be denoted by $\mathcal{M}=\Omega \times \mathcal{K}$. The ``kinetic'' energy of a free particle is $E=D_{2s}\left\| p\right\|^{2s}$ and as a consequence the volume of the set 
\begin{displaymath}
\mathcal{A}=\{(\textbf{x},2\pi \textbf{k})\in \mathcal{M}: \left\| k\right\|\leq \mathcal{E}_{\textrm{max}}^{\frac{1}{2s}}/2\pi \} 
\end{displaymath}
is given by
\begin{eqnarray}
\textrm{Vol}(\mathcal{A})&=& \int_{\Omega}\int_{\left\| k\right\|\leq \mathcal{E}_{\textrm{max}}^{\frac{1}{2s}}/2\pi} dk \, dx\non \\
&=& \frac{1}{(2\pi)^d} \frac{|\Omega| |A_{d-1,2s}|}{d} \mathcal{E}_{\textrm{max}}^{\frac{d}{2s}}.
\label{sec3 : eq3}
\end{eqnarray}
In the case of a hypercube $\Omega=\Gamma_d$ and using (\ref{sec2 : eq10a}) with $\mathcal{E}_{\textrm{max}}=\mathcal{E}_{N}$ we learn that
\begin{displaymath}
\textrm{Vol}(\mathcal{A})=\mathcal{N}+o(\mathcal{N}). 
\end{displaymath}
The sum of eigenvalues of the particle with phase space $\mathcal{A}$ is given by
\begin{eqnarray}
S_{\textrm{class.}}(\mathcal{N})&=& \int_{\Omega}\int_{\left\| k\right\|\leq \mathcal{E}_{\textrm{max}}^{\frac{1}{2s}}/2\pi}\left\| 2\pi k\right\|^{2s} dk \, dx\non \\
&=& (2\pi)^{2s}|\Omega||A_{d-1,2s}| \int_{\mathbb{R}} \chi(0\leq r\leq \mathcal{E}^{1/2s}_{\textrm{max}}/2\pi) r^{d-1+2s} dr \non \\
&=& \frac{1}{(2\pi)^d} \frac{|\Omega||A_{d-1,2s}|}{(d+2s)}\mathcal{E}^{1+\frac{d}{2s}}_{\textrm{max}}
\label{sec3 : eq4} 
\end{eqnarray}
where $\mathcal{E}_{\textrm{max}}$ is taken to be the solution to (\ref{sec3 : eq3}) with $\textrm{Vol}(\mathcal{A})=\mathcal{N}$. 
\\
\textbf{Remarks.} 
\begin{enumerate}
\item Result \rf{sec3 : eq4} is in perfect agreement with the one derived from combining \rf{sec2 : eq13}  with \rf{sec2 : eq10a} to express the sum of eigenvalues in terms of the cut off value $\mathcal{E}_{max}$. 
\item For a tiling domain with $|\Omega'|=\lambda^d |\Omega|$ the eigenvalues scale like \rf{sec2 : eq5c} and the semi-classical sum $S_{\textrm{class.,  P\'{o}lya}}(\mathcal{N})$ scales according to
\beqr
S_{\textrm{class.,  P\'{o}lya}}(\mathcal{N})=\frac{1}{\lambda^{2s}}S_{\textrm{class.,  Weyl}}(\mathcal{N}) \quad \textrm{where} \quad \lambda=\left(\frac{d}{d+2s}\right)^{\frac{1}{2s}\left(1+\frac{d}{2s}\right)}.
\label{sec3 : eq4a}
\feqr
\end{enumerate}
\par Now consider the more realistic case of a moving particle under the influence of a negative valued potential $V\in L^{1+\frac{d}{2s}}(\mathbb{R}^d)$. To estimate the sum of absolute values of the bound states we use the semi-classical approximation inspired from the previous calculation. Thus we have
\begin{eqnarray}
S_{\textrm{class.}}(\mathcal{V})&=& \sum_{n=1}^{N}|\mathcal{E}_n|=\int_{\mathbb{R}^d}\int_{\left\| 2\pi k\right\|^{2s}- \mathcal{V}\leq 0}(-\left\| 2\pi k\right\|^{2s}+\mathcal{V}(x))dk\, dx, \quad \mathcal{V}=\frac{V}{D_{2s}} \non \\
&=& \frac{1}{(2\pi)^d} \int_{\mathbb{R}^d} \chi(\{0\leq \left\| \tilde{k}\right\|^{2s}\leq 1\})(1-\left\| \tilde{k}\right\|^{2s}) d\tilde{k} \int_{\mathbb{R}^d}\mathcal{V}(x)^{1+\frac{d}{2s}} dx\non\\
&=& \frac{1}{(2\pi)^d} |A_{d-1,2s}|\int_{\mathbb{R}} \chi(\{0\leq r\leq 1\})(1-r^{2s}) r^{d-1} dr \int_{\mathbb{R}^d}\mathcal{V}(x)^{1+\frac{d}{2s}} dx\non\\
&=& \frac{1}{(2\pi)^d} \frac{2s}{d+2s}|A_{d-1,2s}|\int_{\mathbb{R}^d}\mathcal{V}(x)^{1+\frac{d}{2s}}dx.
\label{sec3 : eq5} 
\end{eqnarray}
In deriving \rf{sec3 : eq4} and \rf{sec3 : eq5} we used the coordinate transformations
\beqr
x_1&=&r(\cos\theta_1)^{\frac{1}{s}} \non \\
x_2&=&r(\sin \theta_1\cos \theta_2)^{\frac{1}{s}} \non \\
&\vdots& \non \\
x_{d-1}&=&r(\sin \theta_1\cdots \sin \theta_{d-2}\cos \theta_{d-1})^{\frac{1}{s}} \non \\
x_{d}&=&r(\sin \theta_1\cdots \sin \theta_{d-2}\sin \theta_{d-1})^{\frac{1}{s}}, \quad \theta_k\in (0,\frac{\pi}{2}) \,\, \forall k=1,\cdots,d
\label{sec3 : eq6}
\feqr 
with Jacobian determinant
\beqr
J(r,\theta_1,\cdots, \theta_{d-1})=\frac{1}{s^{d-1}}r^{d-1}(\cos \theta_{d-1} \sin \theta_{d-1})^{\frac{1}{s}-1}\prod_{k=1}^{d-2} (\cos \theta_k)^{\frac{1}{s}-1}(\sin \theta_k)^{\frac{k}{s}-1}.
\label{c}
\feqr
The volume of the $2s$-deformed hypersphere of radius $R$ is given by
\beqr
\textrm{Vol}(\Omega_{hyp.})&=& 2^d \frac{1}{s^{d-1}} \frac{R^d}{d} \frac{1}{2^{d-1}}\prod_{k=1}^{d-1}B(\frac{1}{2s},\frac{k}{2s}) 
= R^d \frac{1}{s^{d}} \frac{(\Gamma(\frac{1}{2s}))^d}{\frac{d}{2s}\Gamma(\frac{d}{2s})} \non \\
&=& R^d |A_{d-1,2s}|
\label{sec3 : eq8}
\feqr
where $B$ is the beta function defined by \cite{GR}
\beqr
B(x,y)=2\int_{0}^{\frac{\pi}{2}}\sin^{2x-1}\theta \cos^{2y-1} \theta d\theta, \quad \textrm{Re x, Re y}>0.
\label{sec3 : eq9}
\feqr

%%%%%%%%%%%%%%%%%%%%%%%%%%%%%%%%%%%%%%%%%%%%%%%%%%%%%%%%%%%%%%%%%%%%%%%%%%%%%%%%%%%%%%%%%%%%%%%%%%%%%%%%%%%
%%%%%%%%%%%%%%%%%%%%%%%%%%%%%%%%%%%%%%%%%%%%%%%%%%%%%%%%%%%%%%%%%%%%%%%%%%%%%%%%%%%%%%%%%%%%%%%%%%%%%%%%%%%
\section{Coherent states and the semi-classical approximation for the sum of eigenvalues of the Sch\"{o}dinger operator $D_{2s}\mathcal{L}_{2s}$}

Coherent states have been used extensively to give the leading order semi-classical asymptotics of quantum systems. For reviews see \cite{PE} and references there in.
\begin{definition}
Let $f\in L^2(\mathbb{R}^d)$ be a fixed function with $\left\| f\right\|_2^2=1$. The coherent states associated to f form a family of functions parametrized by $p,y \in \mathbb{R}^d$ such that 
\begin{equation}
G_{y,p}(x)=(\tau_y \circ e_{p}) f(x)=\tau_y \left(e^{\frac{i}{\hbar} \langle p,x\rangle} f(x)\right)=e^{\frac{i}{\hbar} \langle p,x-y\rangle}f(x-y) 
\label{sec4 : eq2}
\end{equation}
where $e_{p}$ is a phase multiplication operator and $\tau_y$ a translation operator.
\label{sec4 : eq1} 
\end{definition}
We will require $f$ to be real or symmetric, i.e. $f(-x)=f(x)$, in what follows. \\

\textbf{\emph{Properties}}
\begin{enumerate}
\item If $f\in L^2(\mathbb{R}^d)$ with $\left\| f\right\|_2^2=1$ then clearly $G_{y,p}(x)\in L^2(\mathbb{R}^d)$ and $\left\| G_{y,p}\right\|_2^2=1$. 
\item If $\psi \in L^2(\mathbb{R}^d)$ its coherent state transform $\tilde{\psi}$ defined by
\begin{equation}
\tilde{\psi}(k,y)=\langle G_{y,k},\psi \rangle=\int_{\mathbb{R}^d} \bar{G}_{y,k}(x)\psi(x)dx \quad \textrm{with} \quad G_{y,k}(x)=e^{2\pi i \langle k,x-y\rangle}f(x-y) 
\label{sec4 : eq3} 
\end{equation}
satisfies
\beqr
\int_{\mathbb{R}^d}\int_{\mathbb{R}^d}|\tilde{\psi}(y,k)|^2 dk dy &:=& \left\| \tilde{\psi}\right\|^2_2 =\left\| \psi\right\|^2_2=\left\| \hat{\psi}\right\|^2_2=1
\label{sec4 : eq3a} \\
\int_{\mathbb{R}^d}\int_{\mathbb{R}^d} \bar{G}_{y,k}(x)G_{y,k}(x') dk\, dy&=&\delta(x-x').
\label{sec4 : eq4} 
\feqr
\end{enumerate}
Relation (\ref{sec4 : eq4}) is interpreted as a weak integral just like Parseval's identity. Also the coherent states $G_{y,k}(x)$ are the rescaled version of \rf{sec4 : eq2} by a factor $\hbar^{1/2}$ for both $p,x$ with the substitution $p=2\pi k$. 
\begin{proposition}
Consider the $L^2(\mathbb{R}^d)$ normalized coherent states
\begin{equation}
G_{y,p}(x)=\frac{1}{\left(2\hbar^{1/2}\Gamma(1+\frac{1}{2s})\right)^{\frac{d}{2}}} e^{\frac{i}{\hbar} (p,x-y)}e^{-\frac{\left\| x-y\right\|^{2s}}{2\hbar^{s}}}.
\label{sec4 : eq5} 
\end{equation}
 Then
\begin{equation}
\lim_{\hbar \rightarrow 0}\langle G_{y,p}, (\mathcal{L}_{2s,\hbar}-\mathcal{V}(y))G_{y,p}\rangle=\left\| 2\pi k\right\|^{2s}-\mathcal{V}(y), 
\label{sec4 : eq6} 
\end{equation}
where the potential $\mathcal{V}\in L^{1+\frac{d}{2s}}(\mathbb{R}^d)$.
\end{proposition}
\textbf{Proof.}
We evaluate first the expectation value of the operator $\mathcal{L}_{2s, \hbar}$ using the coherent states (\ref{sec4 : eq5}).
\begin{eqnarray}
\langle G_{y,p}, \mathcal{L}_{2s, \hbar}G_{y,p}\rangle&=& \langle G_{y,p}, \mathcal{F}_{\hbar}^{-1} \hat{g}_{y,p}\rangle, \quad \hat{g}_{y,p}(\tilde{p}):=\left\| \tilde{p}\right\|^{2s} (\mathcal{F} G_{y,p})(\tilde{p}) \non \\
&=&  \langle \mathcal{F}_{\hbar}G_{y,p}, \hat{g}_{y,p}\rangle, \quad \mathcal{F}_{\hbar}^*=\mathcal{F}_{\hbar}^{-1} \non \\
&=& \int_{\mathbb{R}^d}\bar{\hat{G}}_{y,p}(\tilde{p})\left\| \tilde{p} \right\|^{2s}\hat{G}_{y,p}(\tilde{p})d\tilde{p} \non \\
&=& \frac{|2\pi|^{2s}}{(\left(2\Gamma(1+\frac{1}{2s})\right)^d} \int_{\mathbb{R}^d} \left\|k+\hbar^{\frac{1}{2}} u \right\|^{2s} e^{2\pi i(u,x-z)} du\times \non \\
&& \int_{\mathbb{R}^d} \int_{\mathbb{R}^d} e^{-\frac{\left\|x-y \right\|^{2s}}{2}-\frac{\left\|y-z \right\|^{2s}}{2}} dx dz
\label{sec4 : eq7} 
\end{eqnarray}
where the bar denotes complex conjugation, the positions and momenta have been rescaled by a $\hbar^{1/2}$ factor and $p=2\pi k$. Taking the $\hbar \rightarrow 0$ limit of \rf{sec4 : eq7} we obtain
\beqr
\lim_{\hbar\rightarrow 0}\langle G_{y,p}, \mathcal{L}_{2s, \hbar}G_{y,p}\rangle\!\!\!\!&=&\!\!\!\! \frac{\left\|2\pi k\right\|^{2s}}{(\left(2\Gamma(1+\frac{1}{2s})\right)^d} \int_{\mathbb{R}^d}\int_{\mathbb{R}^d} \!\!\left(\int_{\mathbb{R}^d}e^{2\pi i(u,x-z)}du\right)\!\! e^{-\frac{\left\|x-y \right\|^{2s}+\left\|y-z \right\|^{2s}}{2}}dx dz \non \\
\!\!\!\!&=&\!\!\!\! \frac{\left\|2\pi k\right\|^{2s}}{\left(2\Gamma(1+\frac{1}{2s})\right)^d} \int_{\mathbb{R}^d}\left(\int_{\mathbb{R}^d}\delta(x-z)e^{-\frac{\left\|x-y\right\|^{2s}}{2}} dx\right) e^{-\frac{\left\|y-z\right\|^{2s}}{2}} dz\non \\
\!\!\!\!&=&\!\!\!\! \frac{\left\|2\pi k\right\|^{2s}}{\left(2\Gamma(1+\frac{1}{2s})\right)^d} \int_{\mathbb{R}^d}e^{-\left\|y-z\right\|^{2s}} dz \non \\
\!\!\!\!&=&\!\!\!\! \left\|2\pi k\right\|^{2s}.
\label{sec4 : eq7a}
\feqr
The expectation value of the potential taking into account property (1) is
\begin{equation}
\langle G_{y,p}, \mathcal{V}(y)G_{y,p}\rangle=\mathcal{V}(y)\left\|G_{y,p}\right\|_2^2=\mathcal{V}(y). 
\label{sec4 : eq8a} 
\end{equation}
\hfill\(\Box\) \\
\textbf{Remarks.}
\begin{enumerate}
\item The limit \rf{sec4 : eq7a} can also be derived if one uses the non-unitary Fourier transformation together with a factor modification of Parseval's identity. 
\item In the Gaussian case, $2s=2$,  the normalized function 
\beqr
f(x-y)=\frac{1}{(\pi \hbar)^{\frac{d}{4}}}e^{-\frac{\left\| x-y\right\|^{2}}{2\hbar}}
\label{sec4 : eq8b}
\feqr
is recognized to be the ground state of the d-dimensional isotropic oscillator ($\omega=1$) which minimizes the Heisenberg's uncertainty principle. Moreover from \rf{sec4 : eq7} one can recover the result
\begin{equation}
\langle G_{y,p}, -\hbar^2 \Delta G_{y,p}\rangle=\left\|2\pi k\right\|^{2}_{2}+\frac{\hbar d}{2}.
\label{sec4 : eq9} 
\end{equation}
\end{enumerate}
\begin{theorem}
\label{sec4 : eq9a1}
The semi-classical sum for the moments of eigenvalues of the $Sch\ddot{o}dinger$ operator $\mathcal{H}_{2s}=\mathcal{L}_{2s, \hbar}-\mathcal{V}$ with $\mathcal{V}\in L^{\gamma+\frac{d}{2s}}(\mathbb{R}^d)$, in the $\mathbb{R}^d\times \mathbb{R}^d$ phase space, satisfies
\begin{equation}
\sum_{n}^{class.}|\mathcal{E}_n|^{\gamma}=  \mathcal{C}^{\textrm{class.}}_{2s,\gamma,d}\left\|\mathcal{V}\right\|^{\gamma+\frac{d}{2s}}_{\gamma+\frac{d}{2s}}<\infty, \quad \mathcal{V}\in L^{\gamma+\frac{d}{2s}}(\mathbb{R}^d), \quad \gamma\geq 0
\label{sec4 : eq9a} 
\end{equation}
where
\begin{equation}
 \mathcal{C}^{\textrm{class.}}_{2s,\gamma,d}=\frac{1}{(2\pi)^d}\left(2\Gamma(1+\frac{1}{2s})\right)^d\frac{\Gamma(1+\gamma)}{\Gamma(1+\gamma+\frac{d}{2s})}.
\label{sec4 :eq9b} 
\end{equation}
\end{theorem}
\textbf{Proof.}
Defining the semi-classical trace using \rf{sec4 : eq7a} and \rf{sec4 : eq8a} by
\beqr
\textrm{Tr}(\mathcal{H}_{2s})=\frac{1}{(2\pi)^d}\lim_{\hbar \rightarrow 0} \int_{\mathbb R^d}\int_{\mathbb R^d} \langle G_{x,p}, (\mathcal{L}_{2s, \hbar}- \mathcal{V}(x)) G_{x,p}\rangle dx dp
\label{sec4 :eq9c}
\feqr
 one has
\begin{eqnarray}
\textrm{Tr}(|\mathcal{H}_{2s}|^{\gamma})\!\!\!\!\!&=&\!\!\!\!\! \frac{1}{(2\pi)^d}\int_{\mathbb{R}^d}\int_{\left\|  p\right\|^{2s}\leq \mathcal{V}(x)} (-\left\|  p\right\|^{2s}+\mathcal{V}(x))^{\gamma} dp\, dx \non \\
\!\!\!\!\!&=&\!\!\!\!\! \frac{1}{(2\pi)^d}|A_{d-1,2s}| \!\!\int_{\mathbb{R}^d}\mathcal{V}(x)^{\gamma+\frac{d}{2s}}dx\!\! \int_{\mathbb{R}^d} \!\!\!\chi(\{0\leq r\leq 1\})(1-r^{2s})^{\gamma} r^{d-1}dr \non \\
\!\!\!\!\!&=&\!\!\!\!\! \frac{1}{(2\pi)^d 2s}|A_{d-1,2s}| B\left(\frac{d}{2s},\gamma+1\right)\int_{\mathbb{R}^d}\mathcal{V}(x)^{\gamma+\frac{d}{2s}}dx  \non \\
\!\!\!\!\!&=&\!\!\!\!\! \frac{1}{(2\pi)^d}\left(2\Gamma(1+\frac{1}{2s})\right)^d\frac{\Gamma(1+\gamma)}{\Gamma(1+\gamma+\frac{d}{2s})}\int_{\mathbb{R}^d}\mathcal{V}(x)^{\gamma+\frac{d}{2s}}dx \non \\
\!\!\!\!\!&=&\!\!\!\!\! \mathcal{C}^{\textrm{class}}_{2s,\gamma,d}\left\|\mathcal{V}\right\|^{\gamma+\frac{d}{2s}}_{\gamma+\frac{d}{2s}}<\infty.
\label{sec4 : eq10}
\end{eqnarray}
\hfill\(\Box\)  

%%%%%%%%%%%%%%%%%%%%%%%%%%%%%%%%%%%%%%%%%%%%%%%%%%%%%%%%%%%%%%%%%%%%%%%%%%%%%%%%%%%%%%%%%%%%%%%%%%%%%%%%%%%
%%%%%%%%%%%%%%%%%%%%%%%%%%%%%%%%%%%%%%%%%%%%%%%%%%%%%%%%%%%%%%%%%%%%%%%%%%%%%%%%%%%%%%%%%%%%%%%%%%%%%%%%%%%
%\section{Conclusion}

%%%%%%%%%%%%%%%%%%%%%%%%%%%%%%%%%%%%%%%%%%%%%%%%%%%%%%%%%%%%%%%%%%%%%%%%%%%%%%%%%%%%%%%%%%%%%%%%%%%%%%%%
%%%%%%%%%%%%%%%%%%%%%%%%%%%%%%%%%%%%%%%%%%%%%%%%%%%%%%%%%%%%%%%%%%%%%%%%%%%%%%%%%%%%%%%%%%%%%%%%%%%%%%%%%

\end{document}